\pdfoutput=1
\documentclass[%
 a4paper,
 aip,
 jap,%
 amsmath,amssymb,
reprint,%
]{revtex4-1}

\usepackage{enumitem}
\usepackage{wasysym}
\usepackage[load=abbr]{siunitx}
\usepackage{graphicx}
\usepackage{dcolumn}
\usepackage{bm}

\newcommand{\PRLsep}{\noindent\makebox[\linewidth]{\resizebox{0.3333\linewidth}{1pt}{$\bullet$}}}
\begin{document}

\title{Scientific Opportunies for bERLinPro 2020$+$,\\
Report with Ideas and Conclusions from bERLinProCamp 2019}%

\author{Thorsten Kamps}
\affiliation{Helmholtz-Zentrum Berlin, Berlin, Germany}
\affiliation{Humboldt-Universit\"at zu Berlin, Berlin, Germany}

\author{Michael Abo-Bakr}
\affiliation{Helmholtz-Zentrum Berlin, Berlin, Germany}
\author{Andreas Adelmann}
\affiliation{Paul Scherrer Institut, Villingen, Switzerland}
\author{Kevin Andre}
\affiliation{CERN, Geneva, Switzerland}
\author{Deepa Angal-Kalinin}
\affiliation{ASTEC Daresbury Lab, Warrington, UK}
\author{Felix Armborst}
\affiliation{Helmholtz-Zentrum Berlin, Berlin, Germany}
\author{Andre Arnold}
\affiliation{Helmholtz-Zentrum Dresden-Rossendorf, Dresden, Germany}
\author{Michaela Arnold}
\affiliation{Technische Universit\"at Darmstadt, Darmstadt, Germany}
\author{Raymond Amador}
\affiliation{Humboldt-Universit\"at zu Berlin, Berlin, Germany}

\author{Stephen Benson}
\affiliation{Jefferson Laboratory, Newport News, Virginia, USA}

\author{Yulia Choporova}
\affiliation{Budker Institut for Nuclear Physics, Novosibirsk, Russia}
\author{Illya Drebot}
\affiliation{INFN LASA, Milano, Italy}

\author{Ralph Ernstdorfer}
\affiliation{Fritz-Haber-Institut der Max-Planck-Gesellschaft, Berlin, Germany}
\author{Pavel Evtushenko}
\affiliation{Helmholtz-Zentrum Dresden-Rossendorf, Dresden, Germany}
\author{Kathrin Goldammer}
\affiliation{Reiner Lemoine Institut, Berlin, Germany}

\author{Andreas Jankowiak}
\affiliation{Helmholtz-Zentrum Berlin, Berlin, Germany}
\affiliation{Humboldt-Universit\"at zu Berlin, Berlin, Germany}

\author{Georg Hoffst\"atter}
\affiliation{Cornell University, Ithaca, New York, USA}
\author{Florian Hug}
\affiliation{Johannes-Gutenberg Universit\"at Mainz, Mainz, Germany}
\author{Ji-Gwang Hwang}
\affiliation{Helmholtz-Zentrum Berlin, Berlin, Germany}

\author{Lee Jones}
\affiliation{ASTEC Daresbury Lab, Warrington, UK}

\author{Julius K\"uhn}
\affiliation{Helmholtz-Zentrum Berlin, Berlin, Germany}
\author{Jens Knobloch}
\affiliation{Helmholtz-Zentrum Berlin, Berlin, Germany}
\author{Bettina Kuske}
\affiliation{Helmholtz-Zentrum Berlin, Berlin, Germany}

\author{Andre Lampe}
\affiliation{@andrelampe, Berlin, Germany}


\author{Sonal Mistry}
\affiliation{Helmholtz-Zentrum Berlin, Berlin, Germany}
\author{Tsukasa Miyajima}
\affiliation{High Energy Accelerator Research Organization, KEK, Tsukuba, Japan}

\author{Axel Neumann}
\affiliation{Helmholtz-Zentrum Berlin, Berlin, Germany}
\author{Nora Norvell}
\affiliation{SLAC National Accelerator Laboratory, Menlo Park, California, USA}

\author{Yuriy Petenev}
\affiliation{Helmholtz-Zentrum Berlin, Berlin, Germany}
\author{Gisela P\"oplau}
\affiliation{Universit\"at zu L\"ubeck, L\"ubeck, Germany}

\author{Houjon Qian}
\affiliation{Deutsches Elektronen-Synchrotron, DESY, Zeuthen Site, Zeuthen, Germany}

\author{Hiroshi Sakai}
\affiliation{High Energy Accelerator Research Organization, KEK, Tsukuba, Japan}
\author{Olaf Schwarzkopf}
\affiliation{Helmholtz-Zentrum Berlin, Berlin, Germany}
\author{John Smedley}
\affiliation{Los Alamos National Laboratrory, Los Alamos, New Mexico, USA}

\author{Yegor Tamachevich}
\affiliation{Helmholtz-Zentrum Berlin, Berlin, Germany}
\author{Sebastian Thomas}
\affiliation{Johannes-Gutenberg Universit\"at Mainz, Mainz, Germany}

\author{Jens V\"olker}
\affiliation{Helmholtz-Zentrum Berlin, Berlin, Germany}
\author{Paul Volz}
\affiliation{Helmholtz-Zentrum Berlin, Berlin, Germany}

\author{Erdong Wang}
\affiliation{Brookhaven National Laboratory, Brookhaven, Long Island, USA}
\author{Peter Williams}
\affiliation{ASTEC Daresbury Lab, Warrington, UK}
	
\author{Daniela Zahn}
\affiliation{Fritz-Haber Institut, FHI-MPG, Berlin, Germany}

\date{\today}

\begin{abstract}
The Energy Recovery Linac (ERL) paradigm offers the promise to generate intense
electron beams of superior quality with extremely small six-dimensional phase space for many applications in the physical sciences, materials science, chemistry, health, information technology and security. Helmholtz-Zentrum Berlin
started in 2010 an intensive R\&D programme to address the challenges related to the ERL as driver
for future light sources by setting up the bERLinPro (Berlin ERL Project) ERL with 50 MeV beam
energy and high average current. The project is close to reach its major milestone in 2020, acceleration and recovery of a high brightness electron beam.

The goal of bERLinProCamp 2019 was to discuss scientific opportunities for bERLinPro 2020+. bERLinProCamp 2019 was held on Tue, 17.09.2019 at Helmholtz-Zentrum Berlin, Berlin, Germany.
This paper summarizes the main themes and output of the workshop.
\end{abstract}

\maketitle
\tableofcontents
%

\newpage
\onecolumngrid
\begin{center}
	\begin{figure}[h]
		\includegraphics[width=0.90\textwidth]{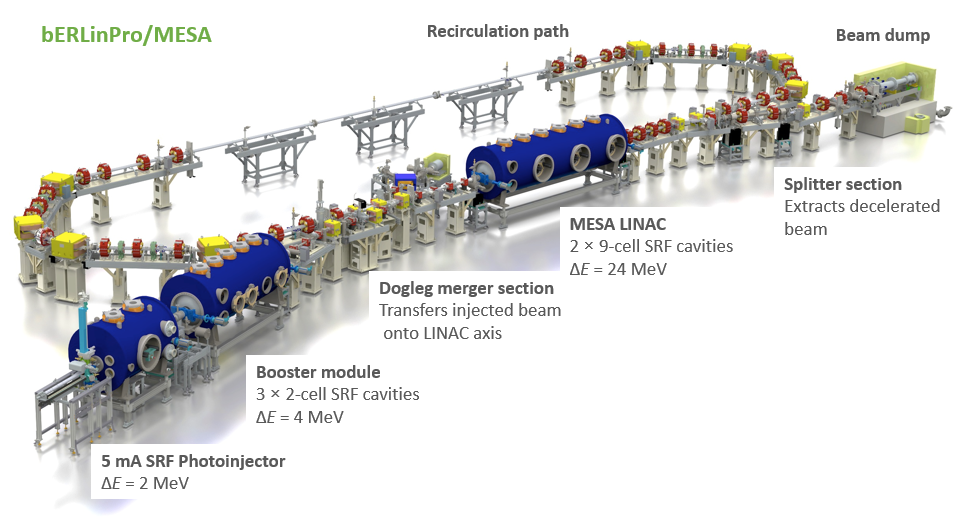}
		\caption{Schematics of the bERLinPro/MESA accelerator complex.}
		\label{fig:bERLinPro_MESA}
	\end{figure}
\end{center}
\twocolumngrid

\section{\label{sec:Intro}Introduction}
The Energy Recovery Linac (ERL) paradigm offers the promise to generate intense
electron beams of superior quality with extremely small six-dimensional phase space for many scientific and technological applications. 

Helmholtz-Zentrum Berlin
started in 2010 an intensive R\&D programme to address the challenges related to the ERL as driver
for future light sources by setting up the bERLinPro~\cite{ref:berlinpro} (Berlin ERL Project) ERL with 50 MeV beam
energy and high average current. The project is close to reach its major milestone in 2020, acceleration and recovery of a high brightness electron beam. This goal will be approached with a base configuration of the bERLinPro accelerator with the MESA linear accelerator module~\cite{ref:MESA,ref:berlinpro_MESA}.

Table~\ref{tab:bERLinProParameter} summarizes the target parameters and their range for the bERLinPro/MESA accelerator.
\begin{table}[hhhh]
	\caption{\label{tab:bERLinProParameter}Preliminary bERLinPro/MESA parameter list.}
	\begin{ruledtabular}
		\begin{tabular}{llcc}
			Parameter & Dimension & Target & Range\\
			\hline
			Beam energy & MeV & 30 & 2 to 45\\
			Beam current & mA &  5 & 0.001 to 5\\
			Repetition rate & MHz & 1300 & 0.1 to 1300\\
			Bunch charge & pC & 77 & 0.1 to 200\\
			Bunch length & fs & 2000 & 10 to 2000\\
			Normalized emittance & mm mrad & 1 & 0.1 to 10\\		
		\end{tabular}
	\end{ruledtabular}
\end{table}
The accelerator complex is shown in Fig.~\ref{fig:bERLinPro_MESA}. The 6 MeV beam energy injection line consits of a single 1.4-cell cavity SRF photoelectron gun with focusing solenoid followed by three 2-cell cavities in the booster module. The beam is then guided on the axis of the MESA accelerator module by a dogleg merger. The MESA accelerator modules houses two 9-cell cavities, which further increase the beam energy to values between 30 and 45 MeV, depending on the performance of the MESA module. Following recirculation and deceleration, the beam is spent at injection energy in a beam dump.

\section{\label{sec:Charge}Workshop charge}
The goal of bERLinProCamp 2019 was to discuss scientific opportunities for bERLinPro/MESA 2020+.
Five main themes were discussed during the workshop:
\begin{itemize}
	\item Injector measurements
	\item Accelerator test facility
	\item Energy doubling and Compton backscattering
	\item Continuous wave superconducting radio-frequency (CW SRF) cavity/module test facility
	\item Multi-color - ultra-fast electron diffraction (UED) plus THz/IR source and applications
\end{itemize}
In the following the main outcome out of each working group is summarized.

\section{Injector measurements}

The goal of the working group was to define some quick wins with existing setup or with relatively minor modifications that can be impactful and achievable within 2-3 years. Previous experience at Daresbury lab indicate that the bulk of work to successfully achieve energy-recovery is in the detailed understanding of the injector.

\begin{enumerate}
	\item Halo investigations: Beam loss of halo particles is a major operational limit for all high power ERL applications. Require control / elimination at $<10^{-7}$ level. bERLinPro to investigate halo generated at injector from any source and benchmark transport of such halo through gun $\rightarrow$ booster $\rightarrow$ merger $\rightarrow$ linac system. Benchmark against simulations as a function of changes to cathode / laser / laser transport / transport element scans. Investigate mitigations, including collimation and other. Requires diagnostics capable of resolving high dynamic range to differentiate between core and halo.
	\item Emittance preservation in merger: Dominant source of beam brightness degradation in high power ERL applications. Utilise diagnostics including transverse deflecting cavity plus spectrometer (TDC/spec) system in straight section to determine 6D phase space. If possible, move this TDC/spec (or install a second one) system to the linac position when linac not present to perform same measurements after merger. Benchmark codes under element scans.
	\item Microbunching investigations: Two contexts – mitigation and enhancement. Mitigation: Utilise TDC/spec diagnostics to study magnitude and evolution as function of cathode choice (if exchange of cathodes feasible), manipulation of laser pulse – perhaps implementation of laser pulse-stacking to enable inter pulse intensity modulation. Enhancement: Investigate feasibility of THz production through deliberate imposition of modulations
\end{enumerate}

\section{Accelerator test facility}
The working group focused on the idea that the various phases of bERLinPro could function as an advanced accelerator test facility. While such facilities are available in the US, Japan and Europe, we are unaware of such a facility in Germany. We focused on two programs that could be instituted at such a facility, and what would be required.  We emphasize that these programs are both very complementary with getting the machine commissioned, and would be good efforts to undertake early in the operation of such a facility.

\subsection{Machine learning}
The first program was machine learning based (ML) optimization of the machine tuning/start-up. To accomplish this, the machine would need to be well instrumented, both in terms of diagnostics to provide feedback and addressable controls systems. The controls should be structured with automated control in mind. Further, this would be most powerful if implemented on a machine with full 6D phase space control. The advantage of doing such a program in the initial stages will be flexibility in the future, allowing the beam to be adjusted quickly to meet the needs of users.

\subsection{Detector testing and calibration}
The second, and related program, is detector and beam diagnostic testing and calibration. While electron beam diagnostics are of interest, the primary focus would be on photodetectors (from IR to X-ray). This would allow the program to benefit from close alignment with the Berlin branch of the German institute for standards and metrology (Physikalisch-Technische Bundesanstalt, PTB). An IR free electron laser (FEL) and a Compton source would be required, as well as precise control of flux, shape, pulse length and pulse repetition frequency. This program could be started alongside the ML program, helping to provide the instrumentation necessary for the ML and eventually benefitting from the flexibility in parameters allowed by the ML program.

\section{Energy doubling and Compton backscattering}
The working group covered two subjects - Energy doubling of the bERLinPro/MESA accelerator and Compton backscattering with the electron beam.

\subsection{Energy doubling}
By changing the length of the recirculation arc it is possible to accelerate the beam in the second pass instead of energy recovering the beam.  This limits the current to several $100~\micA$ but the energy could be as high as 60 MeV.  If the T566 of the lattice can be set correctly it should be possible to cancel out the RF-induced curvature in the longitudinal distribution and get an extremely small energy spread at the full energy.  This would be useful for both Thompson backscattering and Nuclear Physics applications. The advantages and disadvantages are:
\begin{itemize}
	\item[\Large$\smiley$] Can greatly reduce the final energy spread compared to the single acceleration stage. This could be experimentally verfied with the bERLinPro/MESA configuration.
	\item[\Large$\smiley$] Can reach higher photon energies, Compton backscattered photons up to 60 keV.
	\item[\Large$\frownie$] Current limit is half of that for the single acceleration stage.
	\item[\Large$\frownie$] Must alter the downstream arc and add beamline for the higher beam energy beam.
	\item[\Large$\frownie$] Needs full energy beam dump.
\end{itemize}

\subsection{Compton (or Thomson) backscattering}
With a 30 MeV top energy we can reach 20 kV photons from an 800 nm laser. In a basic scheme we can follow an approach implemented at the KEK-ATF~\cite{ref:Compton_KEK}. Furthermore the idea for a zig-zag geometry to obtain head-on collision was discussed. The advantages and disadvantages for Compton backscattering at bERLinPro/MESA are
\begin{itemize}
	\item[\Large$\smiley$] The high repetition rate of the accelerator yields a very high spectral density.
	\item[\Large$\smiley$] We can use a THz undulator or diffraction radiation source to produce THz radiation that can be combined for two color experiments.
	\item[\Large$\smiley$] We can use two stacker cavities at different angles to get two photon energies that could be used for imaging applications~\cite{ref:Drebot}.
	\item[\Large$\smiley$] We can vary the electron bunch length to obtain different bunch lengths from sub-picoseconds to a few picoseconds.
	\item[\Large$\frownie$] Though the flux can be high, the relatively large geometric emittance and low beam energy lead to lower brightness than current synchrotron radiation sources. This means that the ideal applications are those that require short pulses, two colors, or high flux.
	\item[\Large$\frownie$] Very short pulses will have a broader bandwidth.
\end{itemize}

\section{CW SRF cavity/module test facility}
The following scenarios for a CW beam SRF test facility at bERLinPro were discussed:
\begin{itemize}
	\item A full module test probing even for beam break-up instability in case the cavity operating frequency can be adjusted by the return arc for 180 deg of RF phase advance. In that case the module will be in place of the main linac module.
	\item In case a linac module is being operated routinely with bERLinPro, allowing even a parallel usage of that setup with some user experiment, two possible options appear:
	\begin{itemize}
		\item Opposite in the straight of the return arc a cryo-module might be put to be at least tested passively with beam. That would not allow a parallel user experiment and the optics must be checked for the impact on the beam dynamics,
		\item In addition to an e.g. 2 cavity linac module, a short cryo-module might be placed in the linac straight to test single cell prototypes, even of more exotic concepts. This might allow those prototypes to go over the threshold of just being another nice paper design study.
	\end{itemize}
\end{itemize} 
In detail that means:
\begin{itemize}
	\item Depending on frequencies of tested cavities: close the ring and do ERL or not. For closing the ring and operating on different frequencies in ERL mode, a movable return arc would be necessary.
	\item Investigations on higher order modes (also possible if cavities are only passed passively by the beam and are not accelerating). Here we can learn a lot and high beam current would be a unique feature of the bERLinPro test facility.
	\item Test of higher frequency cavities with coatings of different superconductors (e.g. Nb3Sn). These cavities can set path to compact and energy efficient setups in the future but HOMs are even more important to study.
	\item Test exotic cavities with beam (e.g. dual axis). This can boost the knowledge on such cavities and give the chance to develop new technology even though the bigger project might never be realized.
	\item Systematic BBU and instability studies in ERL mode and ERL beam dynamics in general. (needs to close the ring). There are not many ERLs world wide so far so having bERLinPro operational would be a large benefit for ERL community.
\end{itemize}
General remarks:
\begin{itemize}
	\item Using bERLinPro as SRF test facility with beam looks like a good idea but can be rather complicated in details, for example with regards to interfaces and frequencies\cite{ref:Jensen}.
	\item We propose to use the MESA module first and to decide further steps along the findings from module operation. It needs to be stated that any module at a proper operation frequency is necessary to use the ring for (user) experiments, so having a cryomodule on hand is crucial for bERLinPro in the future. If the expeience with the MESA module is positive this could justify to buy such an industry produced model for bERLinPro as a next step.
	\item Integrating different modules is always very expensive due to the cryogenics and vacuum adaptations  (some 100 kEUR), so for considering bERLinPro as a test facility for SRF cavities with beam it could be an appropriate way to integrate a versatile module like HoBiCaT and rather test special cavities than complete dressed modules.
	\item In a later step one could think about building an additional small module to be integrated together with a MESA style module (MiniCaT) for testing exotic cavities.
\end{itemize}

\section{Multi color - UED plus THZ/IR source}
The working group discussed the possibility to use bERLinPro accelerator system and infrastructure to establish a user facility, which would combine Ultrafast Electron Diffraction (UED) probe measurements with THz or IR pump pulses. It might be interesting as well to consider the addition of X-rays for the probe pulses. The X-ray pulses could be supplied by Compton backscattering.

The UED measurements would be done with few (3-4) MeV electron beam energy, and with the repetition rate of up to MHz, which would be a unique capability, very desirable by UED experimentalist, but so far not available worldwide. The repetition rate needs to be flexible as condensed matter pump-probe experiments typically can be performed with kHz repetition rate while gas phase / cw jet experiments can utilize MHz. The energy recovery mode can be considered in the THz generation scheme at high energy with repetition rate of a few MHz

\subsection{THz source}
For the THz/IR pump beams, it is suggested to have two kinds:
\begin{enumerate}[label=(\alph*)]
	\item Broadband - single (few) cycle source(s), and a
	\item Narrowband source with the bandwidth of about 1\% BW or smaller.
\end{enumerate}
Note that option (a) can be realized (although only at kHz repetition rates) with high-power laser systems and is already implemented at the SLAC UED facility (see~\cite{ref:Sie} for first results). So, the key advancement here is the high repetition rate and the use of THz pump/alignment in gas phase studies. For the condensed matter community, option (b) could be more significant.

The broadband - single (few) cycle source(s), which can be based on Coherent Diffraction Radiation (CDR), Coherent Synchrotron Radiation (CSR) or Coherent Edge Radiation (CER), among conventional (well understood) sources. There is also the opportunity to study and use radiators experimentally less tested, but potentially delivering much higher pulse energies. Among such sources are dielectrically loaded waveguides and Coherent Cherenkov Radiation. It would need to be shown that such sources are practical with high repetition rate CW electron beams (with dielectric structures located close to the beam). We expect that such sources would have the bandwidth (appreciable spectral brightness) in the range from $\sim$ 0.1 THz through $\sim$ 2-3 THz. There would be a possibility to compromise between the pulse energy and the spectral range by choosing the bunch charge. Here, a lower bunch charge would provide smaller pulse energy, but would allow shorter bunch length and larger spectral width. The pulse energy could be on the scale of a few $\mu$J. However, it would be important to discuss the necessary pulse energies with the potential UED-THz users.

The narrowband IR/THz source could be realised and an free electron laser (FEL) oscillator operational in CW mode, taking advantage of the CW electron gun and linac availability, which are bERLinPro's specifics. It is suggested that such source should operate covering the wavelength range from $\sim 3 \mu\mathrm{m}$ through $\sim 200 \mu\mathrm{m}$, with out-coupled pulse energy of at least 10 $\mu$J. Again, the wavelength range and the required pulse energy must be discussed with the potential users and reiterated. The optical cavity of the oscillator could most likely be located along the straight section of the return loop.

\subsection{UED source}
In the discussed scheme, at first, a single electron gun would be used to generate very low bunch charge (0.1-1 pC) beam with sufficiently small transfer emittance for UED, and high bunch charge (200 pC or higher) for the generation of THz and IR pump beams. It would need to be shown/checked by beam dynamics modelling that such combination is possible and practical. We noted the risk of increasing the transverse emittance of the low charge UED beam if it would be transported through the merger. Therefore, it was suggested to add an RF kicker close to the booster exit. The idea is that the UED beam would be allowed to go straight i.e. not going to the merger, while the high bunch charge beam receives a kick sending it to the merger. In principle, for such a facility with an average current of about 1 mA, energy recovery is not strictly necessary. If no other future applications of the facility would require the energy recovery more, one could consider reconfiguration of the injector-merger area to simplify the beam optics.


\subsection{Enabling multi-color operation}
The broadband and the narrow band IR-THz sources could be operated simultaneously. For this, the linac needs to accelerate e.g. 10 MHz bunch train for the FEL oscillator interleaved with $\sim$ 1 MHz bunch train for a broadband source. At the exit of the first 180 deg arc, the beam can be split into two by means of a normal conducting RF separator. This would also allow independent longitudinal match (bunch compression) for two beams.

\subsection{Applications}
The following research fields would benefit for a facility combining the THz-IR pump and the UED probe. 

In physics:
\begin{itemize}
\item Material science (new materials / 2D materials / organic crystals / nanostructures): microscopic coupling phenomena, energy flow and conversion,
\item Strongly correlated / quantum materials: non-equilibrium states with specific lattice excitations, light-induced metastable states,
\item Study of heterostructures and operando studies of devices, e.g. transistors, diodes, solar cells, etc.
\item Single-shot experiments of irreversible processes (phase transitions) and transient states like high-density plasma (warm dense matter)
\item In THz pump-probe spectroscopy on doped semiconductor superlattices and graphene is still very strong specially with two synchronized pulses available
\end{itemize}
In chemistry:
\begin{itemize}
\item Molecular dynamics in gas phase, including aligned molecules
\item Radicals and intermediates reaction’s products
\item Operando studies in the field of heterogeneous catalysis
\end{itemize}
In biology:
\begin{itemize}
\item Conformation detection of DNA and RNA for early cancer marker, proteins and water study with THz-IR pump/probe pulses. In a case of just THz radiation there is a room for study with broadband radiation but the source has no unique properties here. In case of UED one doesn’t have non-destructive technique anymore. 
\end{itemize}
Regarding the possible user demand and required beam time: It has been pointed out that the facility at SLAC, has a lot of demand and is very much overbooked. Taking in to account of absence of such facility in Germany (or anywhere) it is easy to imagine that there would be used demand to use this year around - as much as possible.

\section{Conclusions}

The present configuration of the bERLinPro/MESA accelerator opens up many scientific opportunities for the future. Some of those opportunities can already be started today. The main scientific opportunities proposed and discussed during the workshop see bERLinPro 2020$+$ as:
\begin{itemize}
	\item CW SRF beam test facility (CSBTF) for SRF cavities and modules. Here the MESA module integration is a very important first step. Integration of SRF modules into a real accelerator and testing with actual beam is of utmost importance, especially for modules designed for high average current operation. Only with beam higher order modes (HOM) issues can be studied with sufficient depth and only during operation with open ends to the rest of the accelerator will reveal issues from integration in a non-SRF, potentially SRF-"dirty" machine.  
\end{itemize}

\begin{itemize}
	\item Accelerator test facility (ATF) focused on ERL specific questions on injector performance (halo, emittance, microbunching) and multi-turn operation to reach higher beam energies. These injector tests are mandatory for the commissioning of any ERL facility. The possibility to reach higher beam energy with multi-turn operation mode would greatly increase the efficiency of the ERL concept and such the deployability of this scheme.
	\item Multi-color pump/probe user facility (MCPPUF) with many applications driven by THz/IR pump and THz/IR/X-ray/UED probe pulses. For this the X-rays can be generated by Compton backscattering. Unique selling points for such a facility are the multi-color and high repetition rate capabilities. ERL operation would greatly increase the energy efficiency for such a facility
\end{itemize}

Staging these items would mean to first pursue the program outlined with the accelerator test facility (ATF), with the goal to model and understand the bERLinPro accelerator. We will also need to push forward the MESA module integration to learn about the aspects of the CW SRF beam test facility (CSBTF). Armed with this understanding we will be able to setup the multi-color pump/probe user facility (MCPPUF) to better meet user reqirements and be able to offer them a new quality of user beams.

In the short term these items should be discussed further in mini barcamps, each devoted to a single topic.

bERLinPro is the ideal place to pursue a ground-breaking program for accelerator physics R\&D with an CW mA-class SRF energy-recovery linac. Applications will greatly benefit from the flexibility of the facility: in terms of beam parameters and installation options. Furthermore at bERLinPro we can develop capabilities for advanced multi-color radiation generation schemes for THz radiation, X-rays or ultra-fast electron pulses. Delivering this at high repetition rate will enable many scientific applications in physics, chemistry and biology.

\section{The workshop format}
The workshop was organized to follow the barcamp~\cite{ref:barcamp} format (To cite wikipedia: The name barcamp is a playful allusion to the event's origins, with reference to the programmer slang term {\it foobar}). After a quick round for introductions, the participants were asked to develop pitch ideas and briefly present them. After collecting 14 pitches, all individual proposals were discussed and clustered in five working group topics. Due to available space at the workshop venue, all topics could be worked on. After an hour all participants convened again for a plenary session. During this session, the topic speakers presented a summary of what was discussed in the all working groups. The whole event was done in 4h.

The organizers would like to thank all participants for their participation, questions and discussions. We also want to thank HZB for their support and organization of this event.


\PRLsep

\end{document}